\documentclass[aps,pra,showpacs,superscriptaddress,10pt]{revtex4-2}

\usepackage[utf8]{inputenc}
\usepackage[english]{babel}

\usepackage{amssymb}
\usepackage{color}
\usepackage{amsmath}
\usepackage{graphicx}
\usepackage{bm}
\usepackage{braket}

\begin{document}

\title{Scalable quantum circuit design for QFT‐based arithmetic}

\author{Murat Kurt}
\email{kmuratphysics@gmail.com}
\affiliation{Department of Software Engineering, Samsun University, 55420 Samsun, T\"{u}rkiye}

\author{Ayda Kaltehei}
\email{ayda.kaltehei@gmail.com}
\affiliation{Department of Physics, Ondokuz May{\i}s University, 55139 Samsun, T\"{u}rkiye}

\author{Azmi Gen\c{c}ten}
\email{gencten@omu.edu.tr}
\affiliation{Department of Physics, Ondokuz May{\i}s University, 55139 Samsun, T\"{u}rkiye}

\author{Sel\c{c}uk \c{C}akmak}
\email{selcuk.cakmak@samsun.edu.tr}
\affiliation{Department of Software Engineering, Samsun University, 55420 Samsun, T\"{u}rkiye}

\date{\today}

\begin{abstract}
In this research, we create a scalable version of the quantum Fourier transform-based arithmetic circuit to perform addition and subtraction operations on $\mathrm{N}$ $\mathrm{n}$-bit unsigned integers encoded in quantum registers, and it is compatible with $d$-level quantum sources, called qudits. We present qubit- and ququart-based multi-input QFT adders, and we compare and discuss potential benefits such as circuit simplicity and noise sensitivity. The results show that a ququart-based system significantly reduces gate count and improves computational efficiency compared to qubit-based systems. Overall, the findings presented in this study represent a promising step forward in the development of efficient quantum arithmetic circuits, particularly for multi-input operations, with clear advantages for ququart-based systems in reducing gate count, decoherence, and circuit complexity.
\vspace{10pt}
\begin{description}
\item[Keywords]
Quantum arithmetic, Quantum computing, Quantum Fourier transform
\end{description}
\end{abstract}

\maketitle

\section{\label{sec:intro} Introduction}
In quantum computing, the fundamental unit of information is known as a qubit (quantum bit). For multivalued quantum logic ($d>2$), the unit of quantum information is called a qudit~\cite{Bennett00,Keyl02,Wang20}. The unique features of quantum mechanical principles, such as superposition and entanglement, allow for greater computational power and more efficient algorithms~\cite{Dugic02,Harrow17,Montanaro16}. Quantum algorithms used in quantum computing can be divided into several classes, such as quantum search, quantum simulation, and quantum Fourier transform (QFT)-based algorithms~\cite{Adedoyin22}. Quantum Fourier transform is a quantum analogue of classical Fourier transform that can be used to transform a quantum state into superposition of frequency components~\cite{Camps21,Zhou17}.
The QFT is essential for several key quantum algorithms, including phase estimation, Shor’s algorithm for factorization, and Grover’s search algorithm, all of which rely on the QFT~\cite{Adedoyin22,Shor97,Grover96}.

Quantum arithmetic operations can be performed using quantum circuits and algorithms~\cite{Vedral96,Childs10}. Shor developed a quantum algorithm for factoring integers~\cite{Shor97}. Quantum networks for elementary arithmetic operations are presented, and several elementary quantum networks from simple quantum addition to modular exponentiation were constructed by Vedral et al.~\cite{Vedral96}. Quantum carry-save arithmetic was implemented by Gossett~\cite{Gossett98}. A quantum addition based on the ripple-carry approach was presented by Cuccaro et al.~\cite{Cuccaro04}, using a single ancillary qubit. Takahashi and Kunihira conducted several studies related to quantum arithmetic, including Shor’s factoring algorithm~\cite{Takahashi05,Takahashi06,Takahashi08}. A review of quantum arithmetic circuits was later presented by Takahashi~\cite{Takahashi09}. These quantum arithmetic methods utilize multiple $\mathrm{CNOT}$ gates, including the Toffoli gate.

The QFT-based arithmetic adder is first explored by Draper~\cite{Draper00}. Then, algorithm for modular and non-modular arithmetic operations (addition, subtraction, multiplication, division and exponentiation) are widely studied in qubit-based quantum computing~\cite{Draper00,Beauregard03,Maynard14,Perez17,Sahin20,Zhang20,Crimmins24,Pavlidis21,Pachuau22,Paler22,Jakhodia22}. The QFT-based modular arithmetic to Shor's factoring algorithm, illustrating how quantum modular exponentiation can improve computational efficiency for large-number factorization~\cite{Beauregard03}. In recent years, the exploration of QFT-based arithmetic has expanded to include new designs and multi-level (qudit) systems. A quantum integer comparator based on the QFT-based subtractor has been proposed~\cite{Yuan23}. Additionally, modular arithmetic operations are performed on two $n$-bit numbers in qudit-based implementations~\cite{Pavlidis21}.

In this research, we present an extension of the qudit-based QFT-adder, as introduced in~\cite{Pavlidis21}, to perform simultaneous addition and subtraction of $\mathrm{N}$ unsigned integers. This design can be considered a complete version of the qudit-based (multi-level) $\mathrm{n}$-bit $\mathrm{N}$-input QFT-adder. We also discuss the advantages of the $\mathrm{n}$-bit $\mathrm{N}$-input QFT-adder (or subtractor) and compare the quantum circuit designs based on two-level (qubit) and four-level (ququart) systems.

\section{\label{sec:theory} Theory}
In quantum computing, a quantum register is a collection of qubits that together hold a quantum state. This quantum register can be used to encode $\mathrm{n}$-bit classical numbers or superpositions of many such numbers~\cite{Ekert01}. In this work, we consider $\mathrm{n}$-bit classical numbers encoded in quantum registers, and we apply quantum gates to these registers to realize arithmetic operations. During the encoding of the $\mathrm{n}$-bit numbers into quantum registers, the state of the qubits in the quantum register is initially set to $\ket{0}$. Then, quantum $\mathrm{NOT}$ ($\mathrm{X}$) gates are applied to the corresponding qubits to obtain $\ket{0}\rightarrow\ket{1}$ (for converting the relevant bit of number from $0$ to $1$).
This section provides an overview of the theoretical concepts of quantum Fourier transform-based arithmetic.
\subsection{Quantum Fourier Transform}
In quantum computing, the quantum Fourier transform (QFT) offers several advantages, including exponential speedup, parallelism, and reduced circuit depth. The QFT is a quantum version of the discrete Fourier transform that brings the quantum system into a state of superposition and manipulates the amplitudes of a quantum state. It can be expressed as follows~\cite{Pavlidis21}:
\begin{equation}\label{QFT}
\mathrm{{QFT\ket{a}} = \frac{1}{\sqrt{d^n}}\sum_{k=0}^{{d^n}-1}e^{2\pi i.ak/{d^n}}\ket{k}}
\end{equation}
where $\ket{a}$ is the quantum state in computational basis, $n$ is the number of the qudits in $\ket{a}=\ket{a_0 a_1 a_2 \ldots a_{n-1}}$, and $\ket{k}=\ket{k_0} \otimes \ket{k_1} \otimes \cdots \otimes \ket{k_n}$ is the superposition state to be phase encoded ~\cite{Perez17}. It acts on the quantum state as follows;

\begin{equation} \label{QFT_EXPANDED}
\begin{aligned}
\mathrm{QFT} \ket{a}=&\frac{1}{\sqrt{d^n}}\bigotimes_{l=1}^{n} \left[\ket{0}+e^{2\pi ia{d^-l}}\ket{1}+e^{4\pi ia{d^-l}}\ket{2} +\cdots+e^{(d-1)2\pi ia{d^-l}}\ket{d-1}\right]
\end{aligned}
\end{equation}

Since the operators in all processes are Hermitian ($\mathrm{QFT}^{-1}=\mathrm{QFT}^{\dagger}$), the inverse quantum Fourier transform ($\mathrm{IQFT}$) is easily written as;
\begin{equation}\label{IQFT}
\mathrm{{IQFT\ket{k}} = \frac{1}{\sqrt{d^n}}\sum_{a=0}^{{d^n}-1}e^{-2\pi i.ak/{d^n}}\ket{a}}.
\end{equation}
The quantum Fourier transform can be applied on the quantum computing system utilizing the Hadamard ($\mathrm{H}$), controlled-phase-shift ($\mathrm{CP_d(\theta)}$) and SWAP quantum logic gates. The generic expressions of the $\mathrm{H}$ and $\mathrm{CP_d(\theta)}$ gates for $d$-level quantum systems are given as follows:
\begin{equation}\label{hadamard}
\mathrm{{H_d}=\frac{1}{\sqrt{N}}\sum_{j=0}^{d-1}\sum_{m=0}^{d-1}e^{2\pi i(jm)/d}}\ket{j}\ket{m},
\end{equation}
\begin{equation}\label{cp_shift}
\mathrm{{CP_d{(\theta_k)}}=\sum_{j=0}^{d-1}\sum_{m=0}^{d-1}e^{i {\theta_k} (j.m)}}\ket{j}\bra{j}\otimes\ket{m}\bra{m},
\end{equation}
respectively~\cite{Pavlidis21}.
\subsection{QFT-Based Arithmetics}
\begin{figure}[!htb]\centering
\includegraphics[width=10.0cm]{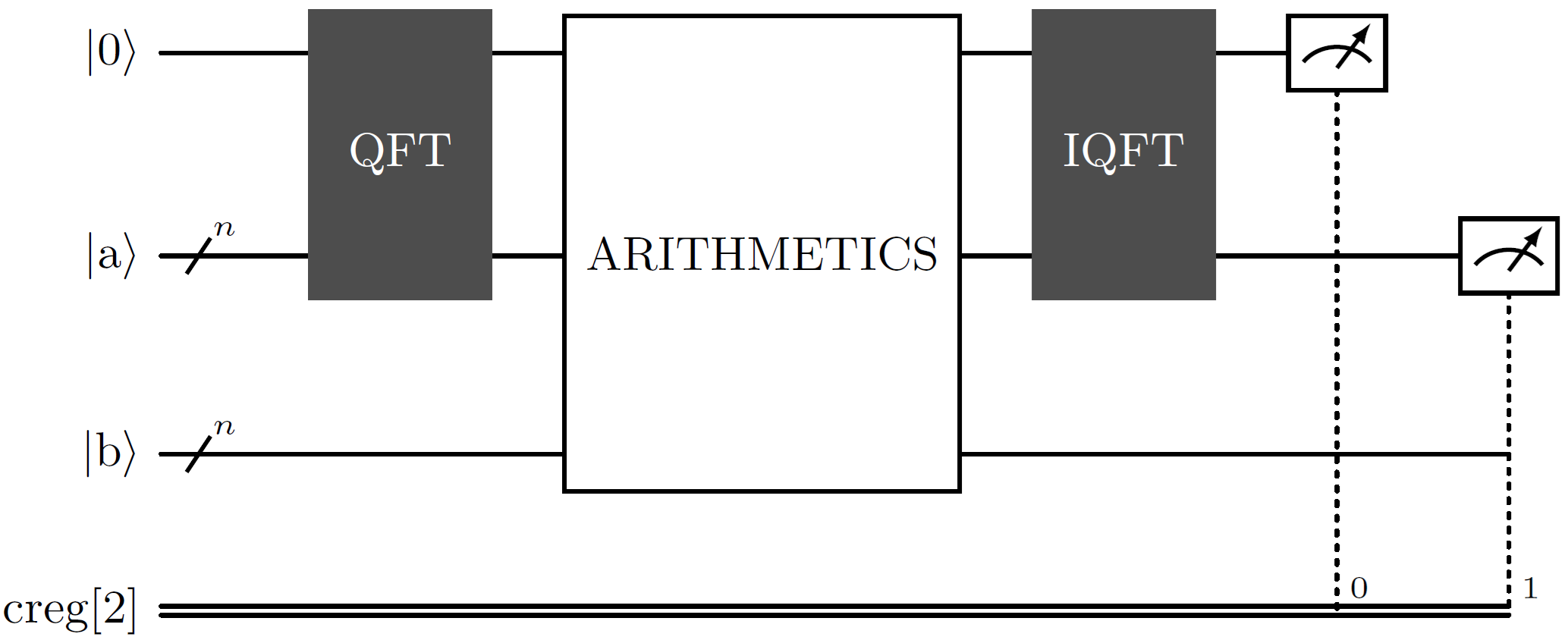}
\caption{\label{BasicQFTadder} The basic circuit design for QFT-based arithmetics. $\ket{a}$ and $\ket{b}$ represents the two $\mathrm{n}$-bits integers are encoded to quantum registers. The measurement outcome holds the result of the aritmetic operation in the classical register (creg).}
\end{figure}

A quantum Fourier transform-based circuit can facilitate simultaneous computations. In QFT-based arithmetic, before applying arithmetic operations, the QFT transforms the initial state into a superposition and encodes the magnitudes (amplitudes) of the numbers into the phases of the states in Fourier space. The generic quantum circuit structure of the QFT-based aritmetics is presented in Fig.~\ref{BasicQFTadder}. Basically, this circuit performs arithmetic operations on two numbers. To prepare for applying the quantum gate for arithmetic operations, the QFT operator is first applied to one of the numbers along with ancillary qudits, transforming the computational basis into the Fourier basis and creating a superposition state. Then, the controlled-phase-shift gates, $CP_d(\theta)$, (in Eq.~\ref{cp_shift}) are used for addition (or subtraction) operation. In the final step, the IQFT is applied to obtain the result of the operation in computational basis. Multiplication and division operations can be performed by iterating through cycles of addition and subtraction.

\section{\label{sec:results} Results}
In this section, we present a scalable quantum circuit design to apply QFT-based arithmetic addition and subtraction on unsigned integers, as well as benchmark the qubit- and ququart-based designs for the same computational output capacity.

\subsection{$\mathrm{n}$-bit $\mathrm{N}$-input QFT-based Adder/Subtractor}

\begin{figure*}[!htb]\centering
\includegraphics[width=16.0cm]{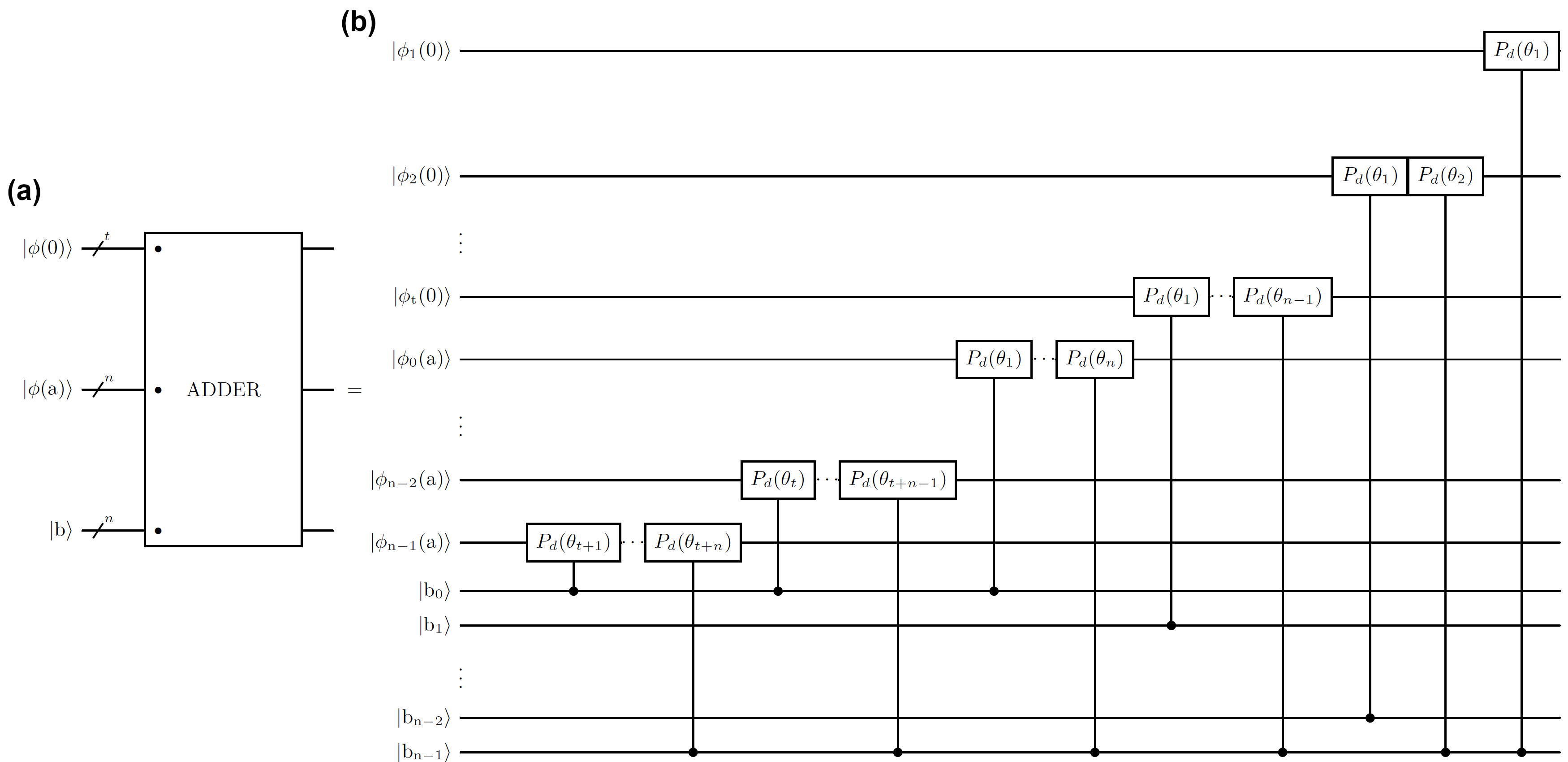}
\caption{\label{qft_adder_part} The gate-like representation of the (non-modular) adder (or subtractor) component consists of three inputs: top, middle, and bottom, which hold quantum registers for the ancillary qubits, the first number, and the second number, respectively (a). The open circuit diagam of generalized $\mathrm{n}$-bit $\mathrm{two}$-input QFT-adder (subtractor) (b). Here, $\ket{\phi(0)}$ and $\ket{\phi(a)}$ are in Fourier base where $\ket{b}$ is in computation base. The first and second numbers are $a$ and $b$, which are initially encoded in quantum registers according to the binary number system, with $a = \{a_0, a_1, \cdots, a_{n-1}\}$ and $b = \{b_0, b_1, \cdots, b_{n-1}\}$.}
\end{figure*}

We introduce the complete version of the QFT-adder (subtractor) circuit, which is also scalable to perform arithmetic operations on $\mathrm{N}$ integers where each integer consists of $\mathrm{n}$-bit encoded on the quantum register. We refer for the such circuit as $\mathrm{n}$-bit $\mathrm{N}$-input QFT-adder (or subtractor). Here, input means quantum register. We first construct the generic circuit for realizing the addition and subtraction operations required for the complete design. We take the quantum circuit presented by Pavlidis~\cite{Pavlidis21} and reconstruct it to generalize for making it compatible with the $\mathrm{N}$ integers ($\mathrm{N} \geq 2$). So, we add the scalable ancillary qubits (or qudits) to the circuit as represented in Fig.~\ref{qft_adder_part}(b). The presented design is known as non-modular~\cite{Perez17}. It performs addition or subtraction operations controlled by the phase of the controlled-phase-shift gates, $CP_d(\theta)$ (Eq.~\ref{cp_shift}). For positive phases, it performs addition ($\mathrm{ADD}$), while for negative phases, it performs subtraction ($\mathrm{SUB}$) between two integers ($\mathrm{a}$ and $\mathrm{b}$, $\mathrm{a>=b}$) where each integer consists $n$-bit and represented by the bit array in binary encoding as $a=\{a_0 a_1 \cdots a_{n-1}\}$ and $b=\{b_0 b_1 \cdots b_{n-1}\}$. Additionally, the number of required ancillary qubits ($t$), shown in Fig.~\ref{qft_adder_part}(b), can be calculated using the following equation;
\begin{equation} \label{ancillary_cnt}
\begin{aligned}
\mathrm{t = \log_d N},
\end{aligned}
\end{equation}
where $d$ is the base (corresponding to the count of energy levels in the quantum source), $\mathrm{N}$ is the count of numbers to be computed, $\mathrm{n}$ is the number of bits required to represent each number. The all ancillary qubits or qudits are initially set to $\ket{0}$.

\begin{figure*}[!htb]\centering
\includegraphics[width=14.0cm]{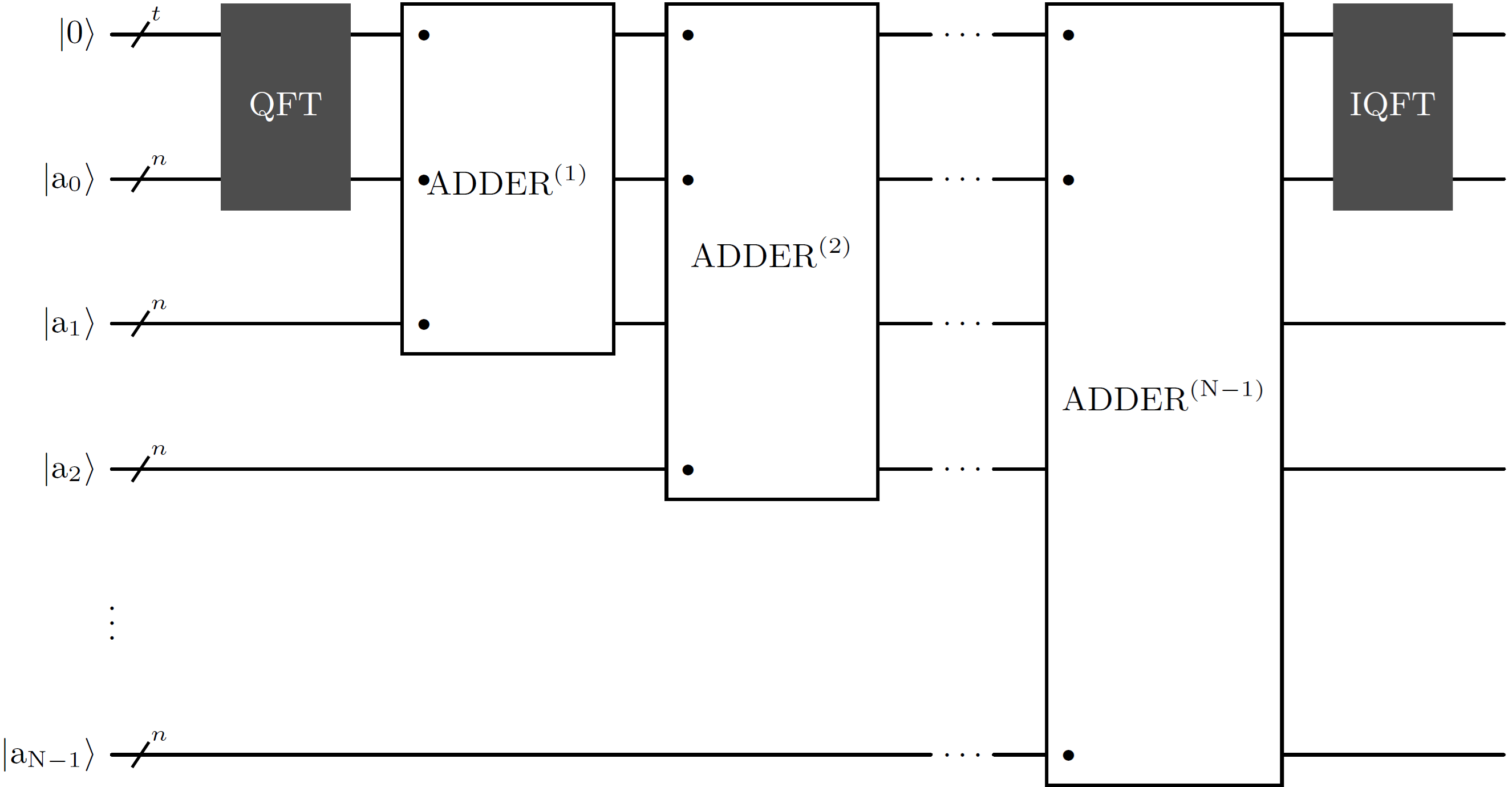}
\caption{\label{complete_version} The generic view of the quantum circuit design uses the $\mathrm{ADDER}$ components shown in Fig.~\ref{qft_adder_part} to design a complete (general) version of the $\mathrm{n}$-bit $\mathrm{N}$-input QFT adder (or subtractor). Here, we remark that each $\mathrm{ADDER}$ component includes three inputs.}
\end{figure*}

The Fig.~\ref{complete_version} shows the complete architecture of the $\mathrm{n}$-bit $\mathrm{N}$-input QFT-adder (or subtractor) utilizing the presented circuits in Fig.~\ref{qft_adder_part}(a) and (b). Here, we note that the input represents the quantum register holding each of the $\mathrm{N}$ numbers (integers). The quantum circuit manipulates the quantum registers and the ancillary qubits throughout following steps: First, the $\mathrm{QFT}$ is applied on the ancillary qudits each is initially set $\ket{0}$ and the quantum register which holds the information of first number $\ket{a_0}$. Thus, ancillary qudits and the quantum register for first number transformed to Fourier base. Second, the adder component, $\mathrm{ADDER^{i}}$, where $i=\{1, 2, 3, \cdots, N-1\}$, is applied via controlled-phase-shift gates, $CP_d(\theta)$, to quantum channels to add numbers sequentially. Please remark here that the each adder part acted on only three input channels (as shown in Fig.~\ref{qft_adder_part}(a)) which one is always ancillary input while other represent two inputs (integers) of N-input. Finally, the inverse quantum Fourier transform ($\mathrm{IQFT}$) converts the quantum state from Fourier basis to computational basis. Thus, the reached final state holds result of complete version of $\mathrm{n}$-bit $\mathrm{N}$-input QFT-adder (or subtractor). Overall, the QFT-based arithmetic addition of $\mathrm{N}$ $\mathrm{n}$-bit integers is applied as follows:

\begin{widetext}
{\setlength{\jot}{8pt}
\begin{align*}\label{QHam}
\ket{\psi_0}=\mathrm{QFT\{\ket{0}^{\otimes t} \otimes \ket{a_0} \}} &=\ket{\phi_0(a_0)} \otimes \ket{\phi_1(a_0)} \otimes \cdots \otimes \ket{\phi_{t+n}(a_0)} \\
\ket{\psi_1}=\mathrm{ADD_{1}} \{ \ket{\psi_0} \otimes \ket{a_1} \} &=\ket{\phi_0(a_0+a_1)} \otimes \ket{\phi_1(a_0+a_1)} \otimes \cdots \otimes \ket{\phi_{t+n}(a_0+a_1)} \\
\ket{\psi_2}=\mathrm{ADD_{2}} \{ \ket{\psi_1} \otimes \ket{a_2} \} &=\ket{\phi_0(a_0+a_1+a_2)} \otimes \ket{\phi_1(a_0+a_1+a_2)} \otimes \cdots \otimes \ket{\phi_{t+n}(a_0+a_1+a_2)} \\
&\vdots\\
\ket{\psi_{N-1}} = \mathrm{ADD_{N-1}} \{\ket{\psi_{N-2}} \otimes \ket{a_{N-1}} \} &=\ket{\phi_{0}(a_0+a_1+\cdots+a_{N-1})} \otimes \cdots \otimes \ket{\phi_{N-1}(a_0+a_1+\cdots+a_{N-1})} \\
\ket{\psi_{N}} = \mathrm{IQFT} \{ \ket{\psi_{N-1}} \} &= \ket{(a_0+a_1+\cdots+a_{N-1})_0} \otimes \cdots \otimes \ket{(a_0+a_1+\cdots+a_{N-1})_{t+n}}
\end{align*}}
\end{widetext}
where $\ket{\psi_{N}}$ represents the addition of the $N$ numbers ($a_0 + a_1 + \cdots + a_{N-1}$) in computational basis. To obtain the result in base-$d$ as a real number, $n + t$ measurements are applied to the quantum registers. The measurement outcome $\ket{(a_0 + a_1 + \cdots + a_{N-1})_{t+n}}$ corresponds to the $(t + n)^{\text{th}}$ bit of the result, which is stored in a classical register (creg) as $\{c_0 c_1 c_2 \cdots c_{t+n}\}_d$ (in base-$d$).

\subsection{Benchmarking the Case Studies}
We execute two cases for $\mathrm{n}$-bit $\mathrm{N}$-input QFT-adder to compare the circuits simplicity and cost of the quantum circuit for qubit-based ($d=2$) and ququart-based ($d=4$) architectures.

\begin{figure*}[!htb]\centering
\includegraphics[width=18.0cm]{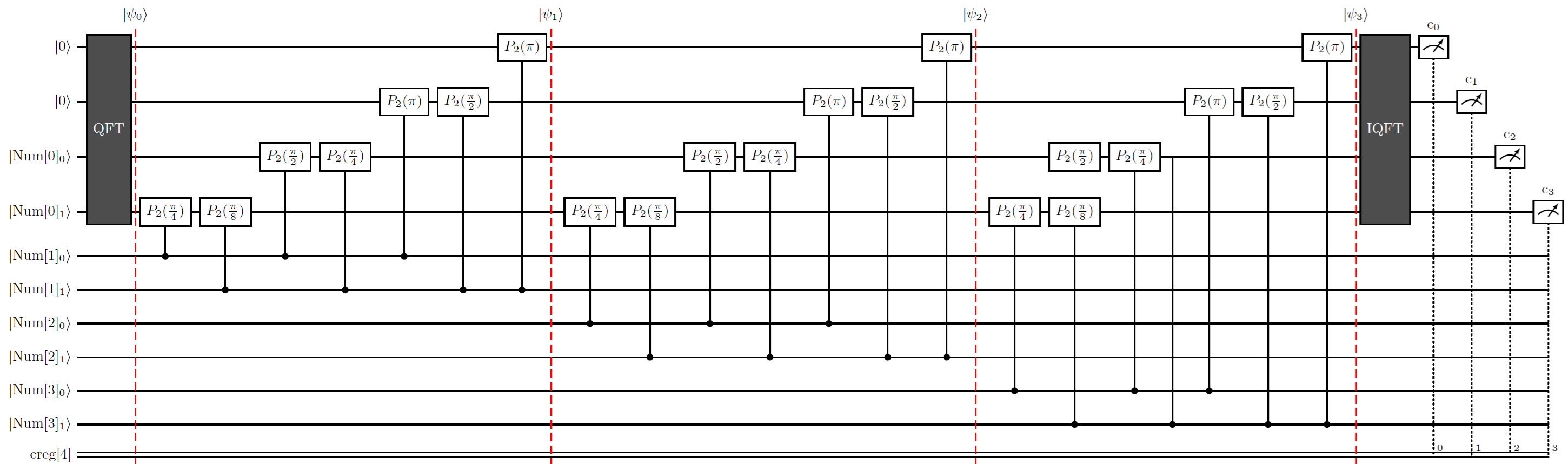}
\caption{\label{qubit_based} The qubit-based ($d=2$) quantum circuit for $\mathrm{two}$-bit $\mathrm{four}$ input QFT-adder. The four two-bit numbers, $\mathrm{Num}[i]$, where $i = \{0, 1, 2, 3\}$, are initially encoded as $\ket{\mathrm{Num}[i]_0, \mathrm{Num}[i]_1}$ in the qubit (quantum) registers. Here, we set the two-bit numbers $\mathrm{Num}[i] = \{3,2,1,2\}$, which are encoded on the qubit states as $3 = \ket{11}$, $2 = \ket{10}$, $1 = \ket{01}$, and $2 = \ket{10}$, respectively. Note that each qubit channel holds one bit of the two-bit integers in base-2. The measurement outcome of the circuit is stored in a 4-bit classical register (creg[4]) in the order $c_0 c_1 c_2 c_3$ in base 2.}
\end{figure*}

\begin{figure}[!htb]\centering
\includegraphics[width=10.0cm]{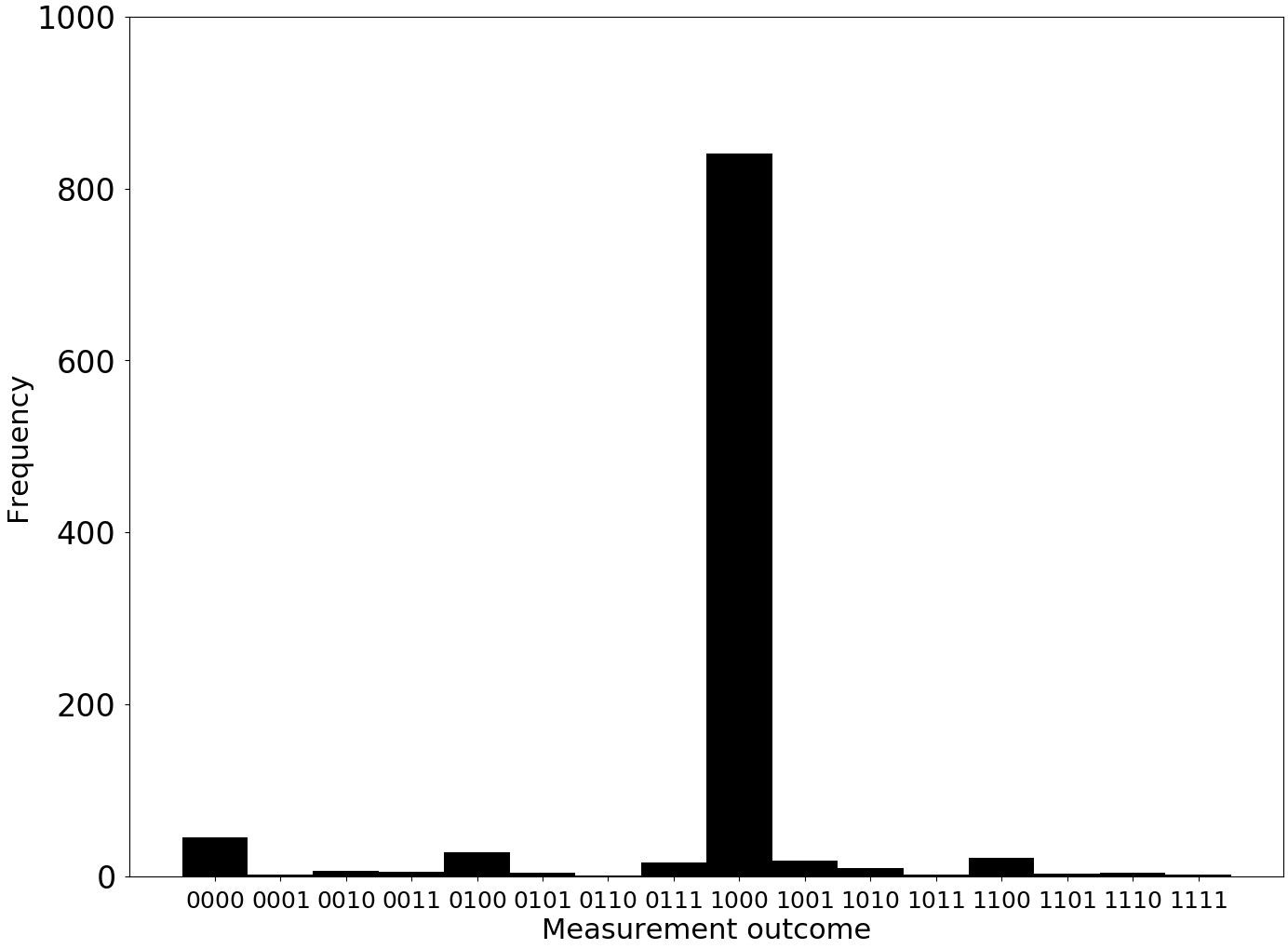}
\caption{\label{measurement} The simulation outcome versus frequency chart for a  $\mathrm{two}$-bit $\mathrm{four}$-input QFT adder. The measurement output of each qubit channel is held in a 4-bit classical register (creg) in the order $c_0 c_1 c_2 c_3$ in binary base (base-2).}
\end{figure}

We first drive the circuit for $\mathrm{n=2}$, $\mathrm{N=4}$ and $d=2$ (qubit logic) that represents the addition of $\mathrm{4}$-numbers each number is $\mathrm{2}$-bit and encoded on qubits shown in Fig.~\ref{qubit_based}. Here, the circuit requires two ancillary qubits, can be calculated by Eq.~\ref{ancillary_cnt}. We execute the presented quantum circuit with IBM Qiskit (adding $\%5$ noise) for the two-bit, four-input QFT-adder to add the four decimal numbers $\{3,2,1,2\}$, where each number is initially encoded in a $\mathrm{2}$-qubit quantum register and represented in binary (base-2) as $11_2$, $10_2$, $01_2$, and $10_2$, respectively. In Fig.~\ref{measurement}, the simulation output shows that the result of the addition operation is $\ket{1000}$. This state represents result of ADD operation of four numbers $3+2+1+2=8$ where binary encoded as $1000_{2}$. Thus, we verify the designed quantum circuit for $\mathrm{2}$-bit $\mathrm{4}$-input QFT-adder.

\begin{figure*}[!htb]\centering
\includegraphics[width=12.0cm]{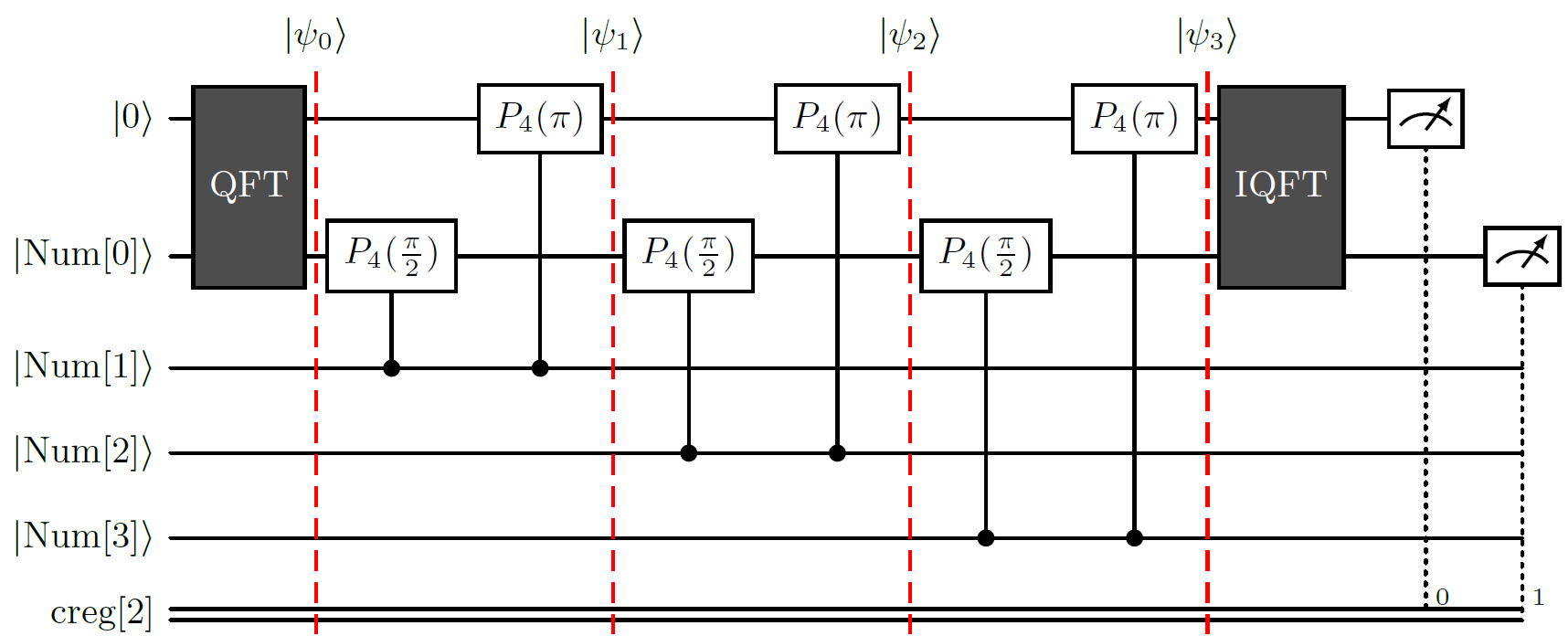}
\caption{\label{qudit_based} The ququart-based ($d=4$) quantum circuit for $\mathrm{2}$-bit $\mathrm{4}$-input QFT-adder. The four two-bit numbers, $\mathrm{Num}[i]$, where $i = \{0, 1, 2, 3\}$, are initially encoded as $\ket{\mathrm{Num}[i]}$ in the ququart registers. Here, we set the two-bit numbers $\mathrm{Num}[i] = \{3,2,1,2\}$, which are encoded on the ququart states as $3 = \ket{3}$, $2 = \ket{2}$, $1 = \ket{1}$, and $2 = \ket{2}$, respectively. Note that each ququart channel holds two-bit number in base-4. The measurement outcome of the circuit is stored in classical register in the order $c_0 c_1$ in base-4.}
\end{figure*}

Second, we construct the quantum circuit for $\mathrm{n=2}$, $\mathrm{N=4}$, and $d=4$ (ququart logic), as shown in Fig.~\ref{qudit_based}. The ququart-based design requires only one ancillary qudit ($t=1$). We consider the same four decimal numbers, given as $\{3,2,1,2\}$, used above, where each number is initially encoded in a ququart ($d=4$) as $3_{4}$, $2_{4}$, $1_{4}$, and $2_{4}$, respectively. As shown in Fig.~\ref{qudit_based}, the number of gates required is significantly reduced compared to the qubit-based circuit version presented in Fig.~\ref{qubit_based}. Since there are no ququart-based (ququart logic) quantum computers or simulators available, we manually perform each gate operation adapted for ququart logic gates~\cite{Wang20}. We obtain the measurement output of the circuit as $\ket{20}$ in the ququart basis. This result corresponds to the same value ($20_{4}=8$) in base-4).

\subsection{Computational Output Capacity}
\begin{figure}[!htb]\centering
\includegraphics[width=8.0cm]{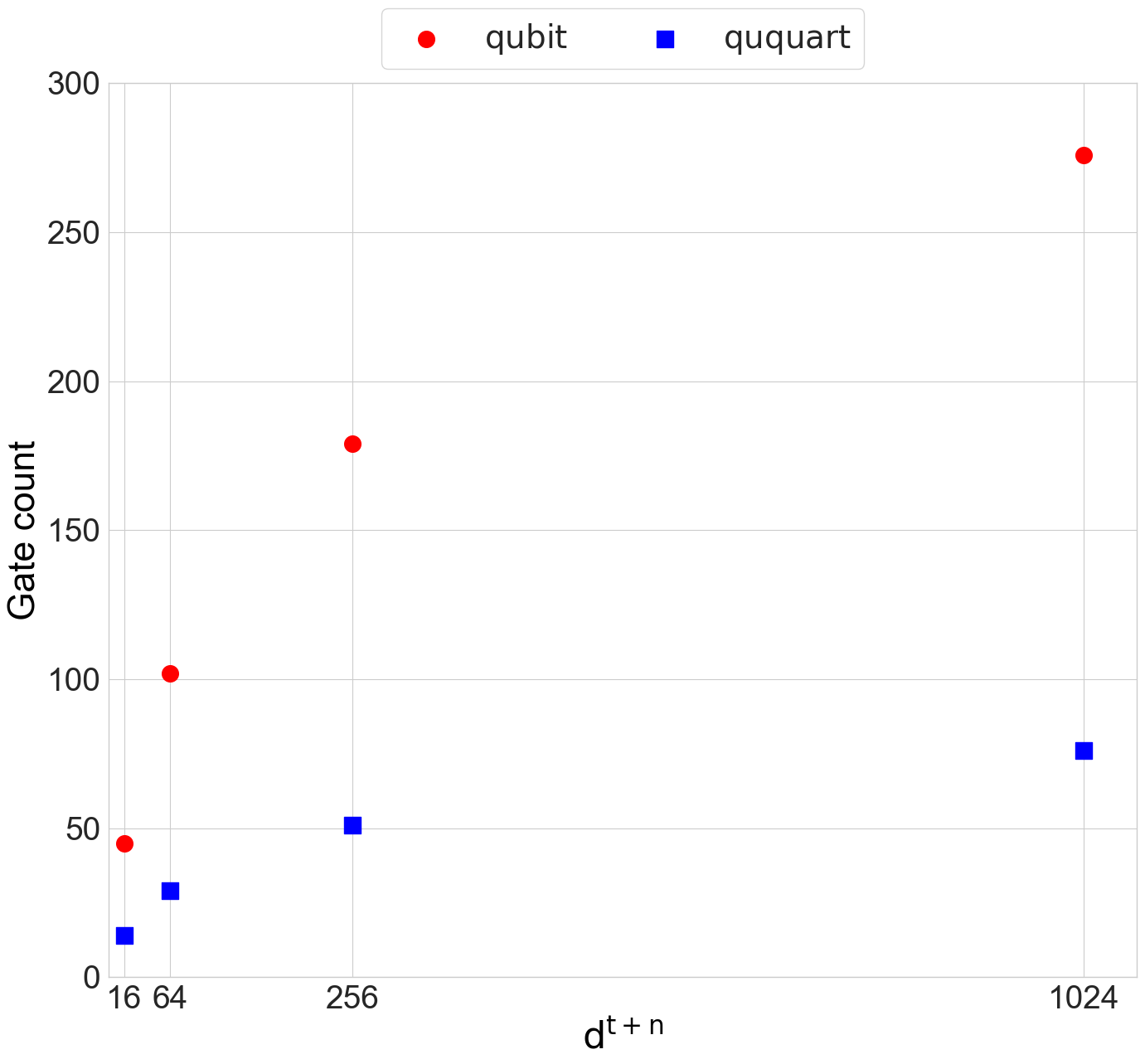}
\caption{\label{gate_count} Gate count versus computational output capacity of $\mathrm{n}$-bit $\mathrm{N}$-input QFT-adder for qubit($d=2$) and ququart($d=4$) based designs.}
\end{figure}

The required gate count for an $\mathrm{n}$-bit, $\mathrm{N}$-input QFT-adder, including all ancillary qubits, is given as follows~\cite{Cakmak24}:
\begin{equation}\label{gate_cnt_even}
\mathrm{Gate\;count}=(N+1)n\left[\frac{(n+1)}{2}+t\right]+t^2+2t+n.
\end{equation}
This equation is valid when $\mathrm{t+n}$ is an even number. If $\mathrm{t+n}$ is an odd number, the generic circuit requires one fewer gate. In Fig.~\ref{gate_count}, we plot the number of gates required to construct an $\mathrm{n}$-bit $\mathrm{N}$-input QFT-based adder versus $d^{\mathrm{t+n}}$ for $d=2$ (qubit-based design) and $d=4$ (ququart-based design). We note that we refer to $d^{\mathrm{t+n}}$ as the computational output capacity of the QFT-based adder (or subtractor). It represents the maximum number that can be defined on the output of the quantum arithmetic circuit, which holds the result of addition or subtraction calculations. It is clear that when comparing the gate counts for obtaining the same computational output capacity, the required gate count significantly decreases when switching the information base (or base dimension) from qubit ($d=2$) to ququart ($d=4$). This simplification benefits quantum circuit design. Additionally, the circuit requires fewer quantum channels, which may help reduce external noise effects. However, we note that the computational cost of constructing elementary gates may increase for higher base dimensions.

\section{\label{sec:conclusion} Conclusions and Discussion}
In conclusion, we designed a scalable QFT-based arithmetic circuit for ADD/SUB operations on a qudit-based quantum source. This architecture enables rapid, large-scale arithmetic operations, as all inputs are processed in parallel. We present a case study that performs the same arithmetic operation on qubit-based and ququart-based designs. When comparing qubit and ququart systems, the results clearly favor the ququart-based design due to its reduced gate complexity. The decrease in the number of gates not only simplifies the quantum circuit but also reduces susceptibility to external noise and decoherence~\cite{Seifert23,Muthukrishnan200,Muthukrishnan2002}. However, it is important to note that while ququart-based systems are theoretically advantageous, the practical implementation of ququart logic gates remains a challenge. Current quantum technologies are primarily optimized for qubits, meaning ququart-based systems may encounter greater practical difficulties in terms of gate construction and error correction.

\begin{acknowledgments}
The authors acknowledge support from the Scientific and Technological Research Council of Turkey (T\"{U}B\.{I}TAK-Grant No. 122F298).
\end{acknowledgments}

\bibliography{references} 
\bibliographystyle{apsrev4-2} 

\end{document}